\documentclass[sigconf]{acmart}

\AtBeginDocument{%
  \providecommand\BibTeX{{%
    \normalfont B\kern-0.5em{\scshape i\kern-0.25em b}\kern-0.8em\TeX}}}

\copyrightyear{2020}
\acmYear{2020}
\setcopyright{rightsretained}
\acmConference[AVI '20]{International Conference on Advanced Visual Interfaces}{September 28-October 2, 2020}{Salerno, Italy}
\acmBooktitle{International Conference on Advanced Visual Interfaces (AVI '20), September 28-October 2, 2020, Salerno, Italy}
\acmDOI{10.1145/3399715.3399744}
\acmISBN{978-1-4503-7535-1/20/09}

\usepackage{enumitem}
\usepackage{color}
\usepackage{graphicx}
\usepackage{subfigure}
\usepackage{epstopdf}
\usepackage{comment}
\usepackage{longtable}
\usepackage{booktabs} % For formal tables
\usepackage{hyperref}

\graphicspath{{images/}}

\begin{document}

\title[BreastScreening]{BreastScreening: On the Use of Multi-Modality in Medical Imaging Diagnosis}

\author{Francisco Maria Calisto}
\email{francisco.calisto@tecnico.ulisboa.pt}
\orcid{0000-0001-8179-7872}
\affiliation{%
\institution{Institute for Systems and Robotics}
\streetaddress{Avenida Rovisco Pais}
\city{Lisbon}
\country{Portugal}
\postcode{1049-001}
}

\author{Nuno Nunes}
\email{nunojnunes@tecnico.ulisboa.pt}
\orcid{0000-0002-2498-0643}
\affiliation{%
\institution{Interactive Technologies Institute}
\streetaddress{Caminho da Penteada}
\city{Funchal}
\state{Madeira}
\country{Portugal}
\postcode{9020-105}
}

\author{Jacinto C. Nascimento}
\email{jan@isr.tecnico.ulisboa.pt}
\orcid{0000-0001-7468-5127}
\affiliation{%
\institution{Institute for Systems and Robotics}
\streetaddress{Avenida Rovisco Pais}
\city{Lisbon}
\country{Portugal}
\postcode{1049-001}
}

\renewcommand{\shortauthors}{Calisto, et al.}

\begin{abstract}

This paper describes the field research, design and comparative deployment of a multimodal medical imaging user interface for breast screening.
The main contributions described here are threefold:
1) The design of an advanced visual interface for {\it multimodal} diagnosis of breast cancer (\hyperlink{https://breastscreening.github.io/}{{\it BreastScreening}});
2) Insights from the field comparison of \hyperlink{https://data.world/mimbcdui-project/single-modality-vs-multi-modality}{Single-Modality vs Multi-Modality} screening of breast cancer diagnosis with 31 clinicians and 566 images; and
3) The visualization of the two main types of breast lesions in the following image modalities:
(i) MammoGraphy (MG) in both 
Craniocaudal (CC) and Mediolateral oblique (MLO) views;
(ii) UltraSound (US); and
(iii) Magnetic Resonance Imaging (MRI).
We summarize our work with recommendations from the radiologists for guiding the future design of medical imaging interfaces.

\end{abstract}




\maketitle

\vfill

\section{Introduction}
\label{sec:sec001}

Breast cancer is the most common cancer in women worldwide~\cite{henriksen2018efficacy}.
Screening plays a fundamental role in the reduction of patient mortality rate.
The most widely employed image modality for breast screening is MammoGraphy (MG).
However, high-risk or dense breast patients require UltraSound (US) or Magnetic Resonance Imaging\footnotemark[1] (MRI) for proper examination~\cite{Maicas2019}.
Therefore, it is quite rare to conduct screening using a \textit{Single-Modality}.

In this paper, we describe the design and comparative testing of \href{https://breastscreening.github.io/}{{\it BreastScreening}} integrating information from several and different image modalities.
We tested the design of \href{https://breastscreening.github.io/}{{\it BreastScreening}} with 31 clinicians noting that the time spent per each image on a \textit{Multi-Modality} strategy is reduced when compared with the \textit{Single-Modality} scenario.
In addition, the lesion classification ({\em e.g.}, \href{https://breast-cancer.ca/bi-rads/}{Breast Imaging Reporting and Data System - BIRADS} ~\cite{SPAK2017179}) is also reduced from our \textit{Multi-Modality} proposed approach.

\subsection{BreastScreening Challenges}

Overall the system involves the  following functionalities:
(1) an interface for identifying (and annotating ground truth) of two types of lesions (i.e., masses and calcifications) across image modalities;
(2) support for categorization of the breast tissues (dense vs non-dense);
(3) a classification (and recommendation) schema for lesion severity using \href{https://breast-cancer.ca/bi-rads/}{BIRADS}~\cite{aghaei2018association, SPAK2017179};
(4) prompt access to clinical co-variables, such as personal and familiar records; and
(5) proper visualizations for a follow-up diagnosis of the patients.

\subsection{Design Process}

The following topics summarize the process we conducted:
(1) findings from a formative study with 31 clinicians, comprising Radiology Room (RR) observations and interviews, which are relevant for both Health Informatics (HI) and Human-Computer Interaction (HCI) fields of research.
This leads us to explore the {\it design goals} (see Section~\ref{sec:sec003});
(2) findings from an evaluation study~\cite{https://doi.org/10.13140/rg.2.2.16566.14403/1} of \href{https://breastscreening.github.io/}{{\it BreastScreening}}, a prototype we developed for the generation of a breast dataset with expert annotations (see Section~\ref{sec:sec004}); and
(3) design recommendations for the use of visualizations to support medical imaging diagnosis (see Sections~\ref{sec:sec004} and \ref{sec:sec005}).

\subsection{Contributions}

In \href{https://breastscreening.github.io/}{{\it BreastScreening}} we provide several new insights, following novel interaction and visualization paradigms~\cite{PAULO2019103316} in the context of breast cancer screening:
$(i)$ multimodal interaction;
$(ii)$ indistinct visualization of cluttered lesions;
$(iii)$ big data management platform; and
$(iv)$ clinicians' multi-screen, multi-environment interaction.
\section{Related Work}
\label{sec:sec002}

This section addresses related work in the HCI field, describing several medical imaging applications.
Our approach covers the limitations of the works following described.
More specifically, we are able to deal with non-homogeneous data.
Comprising multimodal images~\cite{Zhang_2018_CVPR}, classification ({\it i.e.}, BIRADS scores) and annotations ({\it i.e.}, delineation of the lesion contours).

\subsection{Data Visualization}

To our knowledge, few papers~\cite{10.1145/1133265.1133354, 10.1145/2909132.2909248, 10.1145/3206505.3206602} have focused purely on supporting the image search user experience through novel UIs.
These authors described several techniques for presenting all images within a collection in a short time.
Moreover, authors asked users to think and perform browsing an image gallery and selecting an image from the gallery.
These studies, showed us refinement techniques as complements in image systems with relevant user feedback.
However, the presented works are limited to non-clinical users, making it impossible to do a generalization to our research.

\footnotetext[1]{\scriptsize This is the common/current practice in the radiologist services applied in the \href{https://hff.min-saude.pt/}{Hospital Fernando Fonseca} (HFF), Portugal.}

\subsection{Clinical Workflow}

In medical imaging, diagnostic tools enable clinicians to manage patient data, better attend to ongoing tasks and view critical information.
For the diagnostic, understanding the clinical workflow is of chief importance while introducing novel tools and interaction techniques.
Other authors~\cite{10.1145/2685553.2699332, 10.5555/2826165.2826187} present many considerations for collaborative healthcare technology design and discuss the implications of their findings on the current clinical workflow for the development of more effective care interventions.
Supported by the above literature, our goal is to introduce a new tool with several novel interaction techniques, which will improve the final medical imaging diagnosis.

\subsection{Medical Imaging}

From current medical imaging technologies, several issues were identified in the HCI design~\cite{Calisto:2017:TTM:3132272.3134111, calisto2017mimbcdui, Igarashi:2016:IVS:2984511.2984537}.
Some works~\cite{Balducci:2018:BQA:3206505.3206555, Rosado:2015:NFS:2826165.2826213} show the current medical imaging identification techniques for other clinical domains, where most of available systems fail to address the visual nature of the task.
In these two works~\cite{Balducci:2018:BQA:3206505.3206555, Rosado:2015:NFS:2826165.2826213}, the authors create a visual approach to support the \textit{Mental Model} development of the user.
Medical imaging technologies are used to support physicians on the \textit{examination}, \textit{diagnosis}, and (in some cases) \textit{report}~\cite{10.1145/2639189.2639256}.
Others~\cite{10.1145/1385569.1385651, 10.1145/1842993.1843023, Sousa:2017:VVR:3025453.3025566}, study the effectiveness and performance of medical imaging systems, demonstrating how to design a user study for medical imaging experts.
Further, van Schooten et al.~\cite{10.1145/1842993.1843023} measured user performance in terms of time taken and error rate, while interacting with the provided system.
Executing it with several medical users, in this work, the authors show an experiment where their users have similar characteristics as ours.

\subsection{Diagnostic Systems}

Medical imaging has also been extensively studied under the topic of \textit{Computer-Aided Diagnosis (CADx)}, which refers to systems that assist radiologists in image interpretation~\cite{Oram:2014:CDR:2598510.2598585, 10.1145/3359206}.
Wilcox et al.~\cite{Wilcox:2010:DPI:1753326.1753650} propose a design for in-room, patient-centric information displays, based on iterative design with clinicians.
However, these systems are not contemplating the design of an advanced visual interface for multimodal diagnosis on breast cancer disease.
In the above works, we still lack on empirical studies regarding how clinicians can contribute with information contextualization about their clinical workflow, and general medical imaging diagnosis.
Having said that, we also want to add contribution with a study of how medical imaging technologies can play a role in this contextualization.
\section{Design of BreastScreening}
\label{sec:sec003}

The design of \href{https://breastscreening.github.io/}{{\it BreastScreening}} started with a qualitative study to understand radiology practices and workflow in the context of breast screening.
Our study involved 31 clinicians, recruited on a volunteer basis from a large range of clinical scenarios (distinct health institutions in Portugal):
8 clinicians from Hospital Fernando Fonseca; % HFF
12 clinicians from IPO-Lisboa; %IPOL
1 clinician from Hospital de Santa Maria; %HSM
8 clinicians from IPO-Coimbra; %IPOC
1 clinician from Madeira Medical Center; and %MMC
1 clinician from SAMS. %SAMS
Clinicians' experience ranged from 5 - 30 years of medical practice.
The recruited specialists are in advanced career positions and were observed and interviewed in a semi-structured fashion.
Each session took approximately 30 minutes.

\subsection{Standard Clinical Environments}

\href{https://breastscreening.github.io/}{{\it BreastScreening}} works with the standard formats supported by medical imaging~\cite{ng2017technical}, including the MG, US and MRI modalities.
These modalities are available in a standard  Digital Imaging and Communications in Medicine (DICOM) format and supported in Single-Modality by existing systems~\cite{henriksen2018efficacy}. Moreover, most systems are general purpose and do not adapt to specific clinical domains ({\it e.g.}, breast screening). Therefore they do not provide adequate support to the different clinical workflows~\cite{Calisto:2017:TTM:3132272.3134111}.

\subsection{Design Goals}

Combining the clinical context and the technical design challenges lead to a set of design issues, including: \textit{medical imaging structure trade-offs}, \textit{RR temporal awareness}, \textit{image segmentation}~\cite{8736792}, and radiologists system trust.
Based on these, we define five design goals:
\begin{description}
\item[{\it \underline{D}esign around and for \underline{M}edical \underline{I}maging} (DMI):] by taking into account the heterogeneous nature of medical imaging to leverage its contextual richness;
\item[{\it \underline{T}emporal \underline{A}wareness \underline{S}upport} (TAS):] by observing how the radiology workflow events, treatments, and problems progressed over time;
\item[{\it \underline{I}mage \underline{S}egmentation \underline{S}upport} (ISS):] the overview of image details allowing a more accurate diagnostic. Namely, reducing the number of false-positives classification (BIRADS) of the lesion, as well as improving the number of clicks (Section \ref{sec:sec005}) when performing the lesion delineation, {\em i.e.}, segmentation;
\item[{\it\underline{S}everal \underline{M}odalities \underline{S}upport} (SMS):] to enable the view and the process of diagnostic imaging studies, including MG, US and MRI medical imaging modalities;
\item[{\it\underline{G}rowing \underline{T}rust \underline{O}verview} (GTO):] by allowing an efficient triangulation via visualizations, image processing between medical images and available features, {\em i.e.}, annotations of masses and calcifications;
\end{description}

\section{BreastScreening}
\label{sec:sec004}
To validate the proposed design goals, we created \href{https://breastscreening.github.io/}{{\it BreastScreening}}, as a Medical Imaging visualization proof-of-concept to be evaluated in a realistic clinical scenario.
In our design explorations, we sought to integrate several image modalities and visualization to support insight.

%%%%%%%%%%%%%%%%%%%%%%%%%%%%%%%%%%%%%%%%%%%%%%%%%%%
\begin{figure*}[htbp]
\centering
\includegraphics[width=\linewidth]{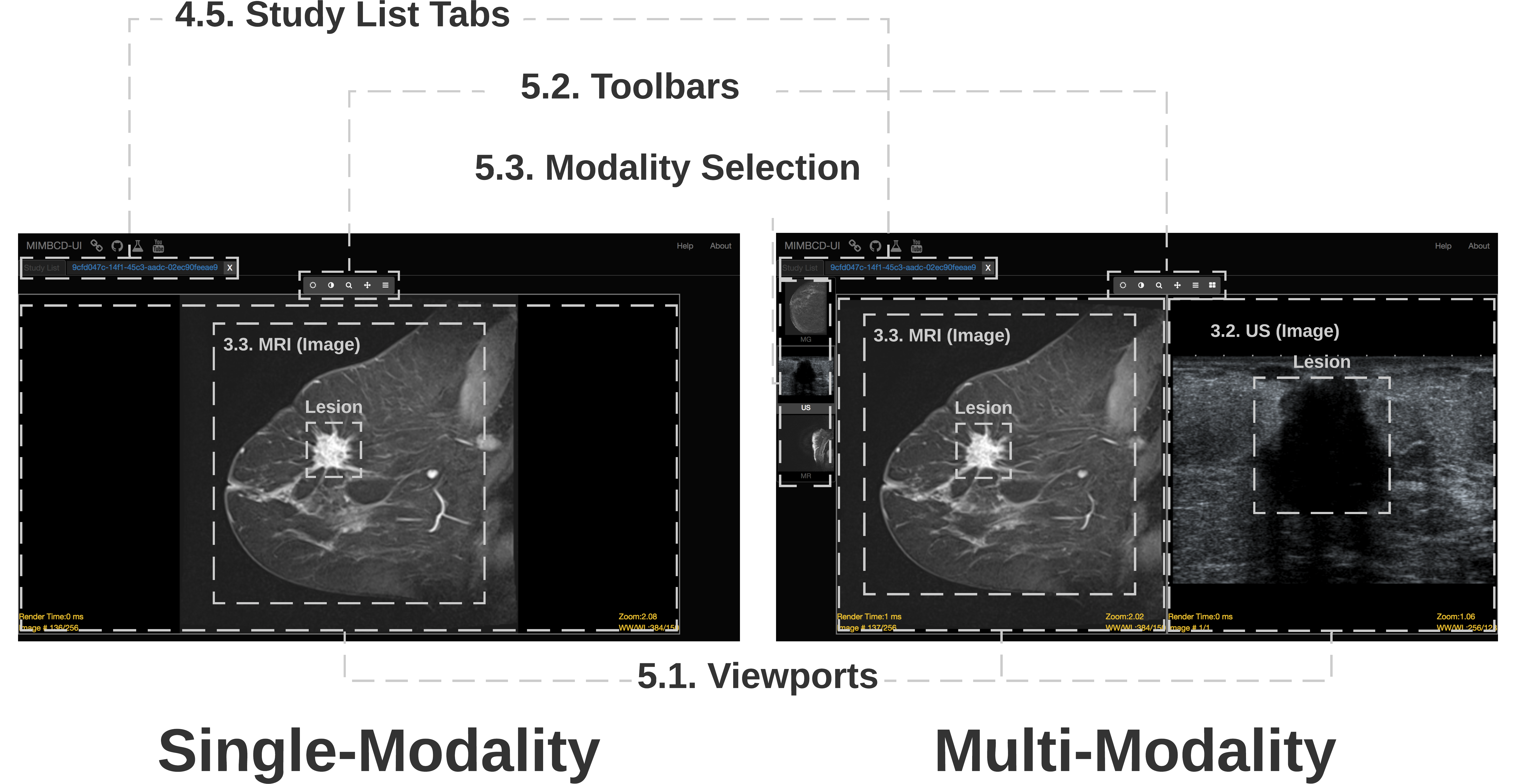}
\caption{\scriptsize \textit{Single-Modality} (left) and \textit{Multi-Modality} (right) Views. The UI components are as follows: \textit{4. List of Patient Views}; and \textit{4.5. Study List Tabs}; as well as \textit{5. Medical Imaging Diagnosis Views}; \textit{5.1. Viewports}; \textit{5.2. Toolbars}; and \textit{5.3. Modality Selection}.}
\label{fig:both_views}
\end{figure*}
%%%%%%%%%%%%%%%%%%%%%%%%%%%%%%%%%%%%%%%%%%%%%%%%%%%

\subsection{User Interface}

The User Interface (UI) consists of two main components:
\textit{4. List of Patient Views}; and
\textit{5. Medical Imaging Diagnosis Views}.
These two main components (Figure \ref{fig:both_views}) are also divided into several sections:
\textit{4.5. Study List Tabs};
\textit{5.1. Viewports};
\textit{5.2. Toolbars}; and
\textit{5.3. Modality Selection}.
Concerning \textit{5. Medical Imaging Diagnosis Views} ({\em Viewports}, {\em Toolbars} and {\em Modality Selection}) this contributes for the temporal awareness (\textit{TAS}).
More specifically, the clinician can probe for lesion patterns~\cite{10.1007/978-3-030-00928-1_62} via the \textit{5.1. Viewports}, processing the image by using the \textit{5.2. Toolbars} features (\textit{GTO}).
The system \textit{5.2. Toolbars} are supporting our image segmentation (\textit{ISS}).
The \textit{5.1. Viewports} are displayed right after the \textit{5.2. Toolbars}, designing around and for medical images (\textit{DMI}) what also improves the temporal awareness (\textit{TAS}) of the task.
On the same time, this design is supporting the way how to interact with several modalities (\textit{SMS}).
Regarding \textit{5.3. Modality Selection}, this allows to the clinician to find more different views (\textit{SMS}) of the same lesion, allowing to perform a better severity classification (Section \ref{sec:sec005}).
Finally, the clinician may look for the lesion shape and contour irregularities (Figure \ref{fig:both_views}) to focus on the segments of the image (\textit{ISS}).
After interacting with the system at the first time, the clinician is able to efficiently process (\textit{ISS}) several images at a same time and use the various given modalities (\textit{SMS}).

\subsection{Implementation}

\href{https://breastscreening.github.io/}{{\it BreastScreening}} was implemented using \href{https://cornerstonejs.org/}{{\it CornerstoneJS}}~\cite{urban2017lesiontracker} with a \href{https://nodejs.org/}{{\it NodeJS}} server.
To populate the system, we selected image sets from HFF patients and upload them into an \href{https://www.orthanc-server.com/}{{\it Orthanc}} server~\cite{Jodogne2018}.
Each patient has three modalities (MG, US and MRI).

The images were pre-processed and anonymized on the \href{https://www.orthanc-server.com/}{{\it Orthanc}} server and then consumed by the \href{https://breastscreening.github.io/}{{\it BreastScreening}} system.
The \href{https://breastscreening.github.io/}{{\it BreastScreening}} core is developed in \href{https://www.w3schools.com/js/}{{\it JavaScript}} with \href{https://www.w3schools.com/jquery/}{{\it jQuery}} for \href{https://www.w3schools.com/html/}{{\it HTML}} document manipulation, event handling and \href{https://github.com/cornerstonejs/dicomParser}{{\it dicomParser}} for parsing DICOM files.
The DICOM files can be loaded by drag-and-drop files into the browser window on the \href{https://www.orthanc-server.com/}{{\it Orthanc}} view.
\section{Results}
\label{sec:sec005}

We conducted an evaluation of \href{https://breastscreening.github.io/}{{\it BreastScreening}} in real-world conditions.
Our goal was to quantitatively and qualitatively assess the proposed design principles and to understand how these principles will play in practice~\cite{10.1145/3027063.3027103}.
We are particularly interested in understanding how
the design goals and challenges (Section~\ref{sec:sec003}) are addressed~\cite{Veeraraghavan2018}.
Ultimately, we are focused on clinicians' opinions how to improve diagnostic reliability.
To accomplish this, the clinicians will have first to deal with:
i) new mechanisms of multi-modal data visualization;
ii) identification and delineation of lesions; and
iii) classification of severity ({\em i.e.} BIRADS).
The experimental setup aimed at testing two conditions:
\textit{Cond. C1} - \textit{Single-Modality}, and
\textit{Cond. C2} - \textit{Multi-Modality}.
For each condition ({\it i.e.}, {\it Single-Modality} or {\it Multi-Modality}) we collected complete imaging exams for three patients (\textit{P1}, \textit{P2} and \textit{P3}) on all possible modalities (MG, US and MRI).
The MG and US comprise a single 2D image ({\em i.e.}, static modality), whilst the MRI~\cite{8759179, SANTIAGO20189} comprises a volume with N slices ({\em i.e.}, dynamic modality~\cite{8296581}).
The exams were previously annotated and classified with a BIRADS severity from an expert doctor who leads the HFF radiology department.

\subsection{Participants}

Our study involved 31 clinicians, recruited on a volunteer basis from a broad range of clinical scenarios, including six different health institutions (two public hospitals, two cancer institutes and two private clinics).
From the demographic questionnaires: 16.10\% of the clinicians have between 31 and 40 years of practical experience (Seniors), 45.20\% have between 11 and 30 years of experience (Middles), 9.70\% have between 6 and 10 years of experience (Juniors), and 29\% have limited experience (Interns).
Interviews were conducted in a semi-structured fashion taking about 30 minutes.
Overall, 17 days were spent on the clinical institutions for the observation process and six months for the classification.

\subsection{Quantitative Analysis}

Four relations\footnotemark[2] emerged from our analysis:
a) differences between \textit{SUS Scores} and \textit{SUS Questions}~\cite{Tyllinen:2016:WNN:2858036.2858570} among clinical experience ({\em i.e.}, \textit{Intern}, \textit{Junior}, \textit{Middle}, and \textit{Senior});
b) the workload measurements of both \textit{Single-Modality} and \textit{Multi-Modality} views;
c) the relation between \textit{Time} and \textit{Number of Clicks}, clustering by Patient ({\em i.e.}, \textit{P1}, \textit{P2} and \textit{P3}).
The expert classification for the patients used in this study are $BIRADS(P1) = 2$, $BIRADS(P2) = 5$ and $BIRADS(P3) = 3$ respectively, for both \textit{Single-Modality} and \textit{Multi-Modality} views; and, d) the distributions of the BIRADS variation (Figure \ref{fig:birads_chart}).

\footnotetext[2]{\scriptsize Available \href{https://github.com/MIMBCD-UI/meta/wiki/Datasets}{{\it datasets}}: \href{https://mimbcd-ui.github.io/dataset-uta4-sus}{{\it usability}} (\href{https://mimbcd-ui.github.io/dataset-uta4-sus}{mimbcd-ui.github.io/dataset-uta4-sus}), \href{https://mimbcd-ui.github.io/dataset-uta4-nasa-tlx}{{\it workload}} (\href{https://mimbcd-ui.github.io/dataset-uta4-nasa-tlx}{mimbcd-ui.github.io/dataset-uta4-nasa-tlx}), \href{https://mimbcd-ui.github.io/dataset-uta4-time}{{\it time}} (\href{https://mimbcd-ui.github.io/dataset-uta4-time}{mimbcd-ui.github.io/dataset-uta4-time}), \href{https://mimbcd-ui.github.io/dataset-uta4-rates}{{\it severity rates}} (\href{https://mimbcd-ui.github.io/dataset-uta4-rates}{mimbcd-ui.github.io/dataset-uta4-rates}), and \href{https://mimbcd-ui.github.io/dataset-uta4-dicom}{{\it images}} (\href{https://mimbcd-ui.github.io/dataset-uta4-dicom}{mimbcd-ui.github.io/dataset-uta4-dicom}).}

\subsubsection{\textit{SUS Scores} vs \textit{SUS Questions}}

The ANOVA test\footnotemark[3]~\cite{Wobbrock:2011:ART:1978942.1978963} yields a significant difference in both \textit{Single-Modality} (F\textsubscript{SM} = 11.79, p\textsubscript{SM} = 0.001 $<$ 0.05) and \textit{Multi-Modality} (F\textsubscript{MM} = 23.31, p\textsubscript{MM} = 0.001 $<$ 0.05) conditions among the various clinical experience of Clinicians.
Participants adopting the \textit{Multi-Modality} (M\textsubscript{MM} = 2.9, SD\textsubscript{MM} = 0.90) condition obtained higher SUS scores than those using the \textit{Single-Modality} (M\textsubscript{SM} = 2.7, SD\textsubscript{SM} = 1.01) condition.

\footnotetext[3]{\scriptsize \textit{N}: the number of users (Clinicians); $F\textsubscript{var}$: the F-test used for comparing the factors of the total deviation per each variable (\textit{var}) categorized by clinical experience; $M\textsubscript{var}$: Mean value of the variable (\textit{var}); $SD\textsubscript{var}$: the Standard Deviation (SD) per each variable (\textit{var}).}

\subsubsection{\textit{Workload}}

The results generated from the NASA-TLX~\cite{10.1145/3290605.3300592} yields a significant main effect for the Physical Demand (F\textsubscript{SM} = 5.81, p\textsubscript{SM} = 0.003 $<$ 0.05) and Temporal Demand (F\textsubscript{SM} = 4.86, p\textsubscript{SM} = 0.009 $<$ 0.05).
On the other hand, the \textit{Multi-Modality} condition indicates that there exists a significant difference among Mental Demand (F\textsubscript{MM} = 3.13, p\textsubscript{MM} = 0.04 $<$ 0.05), Physical Demand (F\textsubscript{MM} = 4.61, p\textsubscript{MM} = 0.009 $<$ 0.05), and Temporal Demand (F\textsubscript{MM} = 9.17, p\textsubscript{MM} = 0.001 $<$ 0.05).
The NASA-TLX yields significant difference among groups for both Effort (F\textsubscript{MM} = 3.74, p\textsubscript{MM} = 0.02 $<$ 0.05) and Frustration (F\textsubscript{MM} = 3.93, p\textsubscript{MM} = 0.01 $<$ 0.05).

\balance

\subsubsection{\textit{Time} vs \textit{Number of Clicks}}

Results showing the amount of \textit{Time} and \textit{Number of Clicks} in each of the 566 images among the three patients are following described.
The ANOVA test shows a non-significant interaction effect over the total \textit{Time} from both \textit{Single-Modality} (F\textsubscript{SM} = 0.68, p\textsubscript{SM} = 0.56 $>$ 0.05) and \textit{Multi-Modality} (F\textsubscript{MM} = 0.28, p\textsubscript{MM} = 0.83 $>$ 0.05) regarding the clinical experience groups.
In addition, our results show a non-significant interaction effect for the total amount of \textit{Number of Clicks} from both \textit{Single-Modality} (F\textsubscript{SM} = 1.76, p\textsubscript{SM} = 0.17 $>$ 0.05) and \textit{Multi-Modality} (F\textsubscript{MM} = 0.57, p\textsubscript{MM} = 0.63 $>$ 0.05).

\subsubsection{\textit{BIRADS Classification}}

\begin{figure}[ht]
\centering
\includegraphics[width=0.40\textwidth]{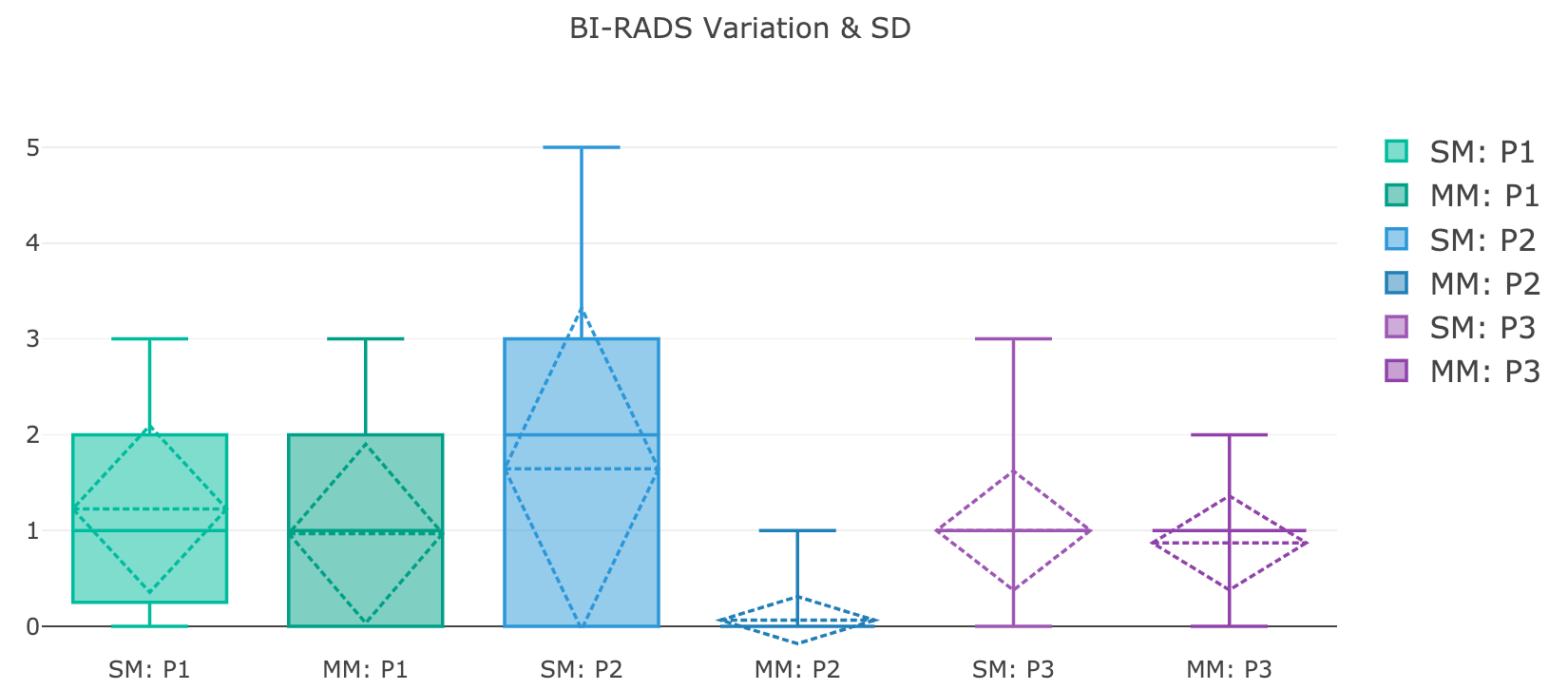}
\caption{\scriptsize BIRADS variations distribution among the 31 clinicians.
We subtract the expert classification from the classification performed by each clinician
(the closer to zero the graph is, the greater the classification is). The \textit{ordinate axis} represent the \textit{BIRADS Values} of a scale between 1 to 5.
The \textit{abscissas axis} represents each Patient ({\em i.e.}, \textit{P1}, \textit{P2} and \textit{P3}) with both \textit{Single-Modality} (SM) and \textit{Multi-Modality} (MM).
The rhombus represents the SD.}
\label{fig:birads_chart}
\end{figure}

The first and second order statistics of the BIRADS classification is shown in Figure~\ref{fig:birads_chart}.
The mean values are referenced to the patient BIRADS (previously performed by the expert), that is, we have (from left to right) the patients \textit{P1}, \textit{P2} and \textit{P3}, with BIRADS\textsubscript{real} = 2, BIRADS\textsubscript{real} = 5 and BIRADS\textsubscript{real} = 3, respectively.
From this figure, it is clear that the \textit{Multi-Modality} performs better, since the most severe BIRADS exhibits the smaller mean and variance ($|$BIRADS\textsubscript{real} - BIRADS\textsubscript{provided}$|$) in the most of the cases. Also note that for the most problematic patient (in this case P2 scored with $BIRADS=5$) the multi-modal largely outperforms the \textit{Single-Modality} setting.

%%%%%%%%%%%%%%%%%%%%%%%%%%%%%%%%%%%%%%%%%%%%%%%%%%%
%%%%%%%%%%%%%%%%%%%%%%%%%%%%%%%%%%%%%%%%%%%%%%%%%%%

\subsection{Qualitative Analysis}

Clinicians were invited to give some feedback about the UI during the open interviews.
We received several positive comments regarding our \href{https://breastscreening.github.io/}{{\it BreastScreening}} system.
At the end, several clinicians (19/31) answered that the assistant will be an asset of an immense importance for the current RR situation:
\textit{``The system will be a great asset for us''} (C6).
Another positive answer was the one related to the frequency of use (28/31) for this new assistant regarding the current system used by the clinicians on the daily practice:
\textit{``I would like to frequently use your system on my daily practice''} (C1).
\section{Conclusion}
\label{sec:sec006}

Medical imaging systems provide a promising but challenging problem for HCI research.
In this paper, we presented field research, design and comparative deployment of a multimodal user interface for breast screening,
\href{https://breastscreening.github.io/}{{\it BreastScreening}} is a proof-of-concept prototype developed to embody the emerging design goals from the underlying clinical context.
Our work and contributions included:
a) identifying the main clinical workflow issues, the interaction cognitive load challenges~\cite{Castner:2018:ONG:3279810.3279845} and the opportunities;
b) establishing a set of design goals for medical imaging design;
c) the design, reflections and in-situ evaluation of \href{https://breastscreening.github.io/}{{\it BreastScreening}} supporting the clinical translation; and
d) the impact evidence of \textit{Multi-Modality} in diagnosing and severity classification of breast lesions with 31 radiologists in six different clinical institutions.
Our results\footnotemark[4] show that the system can lead to more efficient and accurate clinical diagnosis.

\footnotetext[4]{\scriptsize We provide our statistical analysis (\href{https://mimbcd-ui.github.io/statistical-analysis/}{mimbcd-ui.github.io/statistical-analysis}) supporting this study with evidence. Several charts are plotted to help on the visualization of our results.}

\clearpage

\bibliographystyle{format}
\bibliography{bibliography}

\end{document}